\documentclass[onecolumn,showpacs,preprintnumbers,amsmath,amssymb]{revtex4}

\usepackage{epsf}
\usepackage{graphicx}  % Include figure files
\usepackage{dcolumn}   % Align table columns on decimal point
\usepackage{bm}        % bold math

\newcommand{\be}{\begin{equation}}
\newcommand{\en}{\end{equation}}
\newcommand{\bea}{\begin{eqnarray}}
\newcommand{\ena}{\end{eqnarray}}

\begin{document}

\preprint{}

\title{N-dimensional plane symmetric solution with perfect fluid source }

\author{Hongsheng Zhang\footnote{Electronic address: hongsheng@kasi.re.kr} }
 \affiliation{\footnotesize
 Korea Astronomy and Space Science Institute,  Daejeon 305-348, Korea }
 \affiliation{\footnotesize Department of Astronomy, Beijing Normal University,
Beijing 100875, China}
 \author{Hyerim Noh\footnote{Electronic address: hr@kasi.re.kr} }
 \affiliation{\footnotesize
 Korea Astronomy and Space Science Institute,
  Daejeon 305-348, Korea }
 \date{ \today}

\begin{abstract}
 A new class of plane symmetric solution sourced by
 a perfect fluid is found in our recent work. An n-dimensional ($n\geq 4$) global plane symmetric solution of Einstein field equation
  generated by a perfect fluid source is investigated,
 which is the direct generalization of our previous 4-dimensional
 solution.
   One time-like Killing vector and $(n-2)(n-1)/2$ space-like Killing vectors, which
  span a Euclidean group $G_{(n-2)(n-1)/2}$, are
 permitted in this solution. The regions of the parameters constrained by
 weak, strong and dominant energy conditions for the source are
 studied. The boundary condition to match to n-dimensional Taub metric and Minkowski metric
 are analyzed respectively.

\end{abstract}


\maketitle

\section{Introduction}
  A new class of static plane symmetric solution of Einstein field
 equation sourced by a perfect fluid was successfully found in \cite{self}.
 In this solution, the density and pressure behave in a
 non-conventional way: Both the density and pressure are variables
 with respect to spatial position and they are not proportional
 to each other. All the previous solutions have been derived based on the
 suppositions that the density is proportional to the pressure or
 the  density is a constant \cite{exact1, exact2, taub2}.

  This solution  puzzles out
 a well-known problem: the source of Taub solution \cite{taub}. The
 history of searching for the source of Taub solution is a little bit
 long.
  A no-go theorem  was proved in \cite{taub2} more than 50 years ago, which
  said  that a perfect fluid cannot bound a vacuum in a
  space with plane symmetry unless the boundary condition of the
  continuity of the derivatives of the metric tensor is violated.
  This means that there does not exist a matter source which can
  perfectly match to vacuum Taub solution. It seems that the way to find a source for Taub solution is blanked
  off, which significantly lowers the importance of Taub solution.
   A necessary prerequisite for the source matters is imposed in the proof of this
  no-go theorem,  that is,  the pressure
 is positive throughout the slab for the solution. It is a natural requirement for the matters before the
 discovery of cosmic acceleration. Also,  it is shown that
 generally any singularity free source with reflective symmetry for plane symmetric vacuum space
  must violate dominant energy condition (DEC)\cite{dec}. DEC is always
 violated in one branch of the solution in \cite{self}, no matter what values the parameters are taken. Further,
  we have found that there is a configuration of the large branch in
  the class of \cite{self} which can perfectly
  match to vacuum Taub solution, and  it is naturally to be identified as the
  source of Taub solution.

   The properties of geodesics of the solution in \cite{self} are explored in \cite{self2}. It is
   found that this solution can be an appropriate simulation to the field of a uniformly
   accelerated observer in Newton mechanics. A research on plane symmetric solutions in order to find the
   best simulation to general relativity
    of the Newtonian infinite plane was presented in \cite{jone}.
    For some interesting researches on the equivalence between a uniformly accelerating reference frame and
    the gravitational field, see \cite{equi}.
   Though we obtain the exact form of the energy-momentum
   of the perfect fluid sourced solution in \cite{self}, we do not know what it really is. The
   essence of the source is investigated in \cite{self2}. A phantom with dust
   and photon is suggested as the substance of the source matter.

  Higher dimensional theory has a fairly long history since Kaluza-Klein's proposal.
  Recently, it plays more and more important role in high energy physics and cosmology.
       In this letter we will generate the solution with a perfect fluid
  source in higher dimensions. Our investigations are parallel to
  the discussion in \cite{self}. Almost all results can come back to the
  4-dimensional case, except some special situations, which we
  will give proper explanations. The most important reason is that there is a singularity
  in the general results when we apply it to the 4-dimensional
  case. So the 4-dimensional case needs separate investigation, and
  it has been done in our previous work.

 This letter is organized as follows: In the next section we will
 present a n-dimensional solution as a generalision of our previous 4-dimensional solution.
 In section III we investigate the requirements of energy conditions in this space.
 In section IV we study the matching conditions to Minkowski and Taub, respectively. A
    summary is presented in section V.

 \section{the solution}

 We deduce the following n-dimensional ($n\geq 4$) solution of Einstein field equation
 sourced by a perfect fluid in a time orthogonal chat,
 \be
 ds^2=-e^{2az}dt^2+dz^2+e^{2[az+be^{az/(n-3)}]}d\Sigma^2,
  \label{metric}
 \en
 where
 \be
 d\Sigma^2=dx_1^2+dx_2^2+...+dx_{n-2}^2.
 \en
 Clearly the above metric permits 1 time-like Killing fields $\frac{\partial}{\partial
 t}$, which denotes that the present solution is static, and
 $(n-2)(n-1)/2$ space-like Killing fields (including $n-2$
 translational fields
 and $(n-2)(n-3)/2$ rotational ones), which span a Euclidean group $G_{(n-2)(n-1)/2}$.
 The metric (\ref{metric}) is an exact solution of  Einstein field
equation sourced by a  perfect fluid,
 \be
 T=(\rho(z)+p(z))U\otimes U+p(z)g,
 \label{em}
 \en
 where $T$ denotes the energy momentum tensor of the fluid, $U$
 stands for 4-velocity of the fluid, $g$ denotes the metric
 tensor, and,
\be
  \rho=-\frac{a^2}{2}\frac{n-2}{(n-3)^2}\left[(n-3)^2(n-1)+2(n-2)^2be^{az/(n-3)}
  +(n-1)b^2e^{2az/(n-3)}\right],
  \label{rho}
 \en
 \be
  p=\frac{a^2}{2}\frac{n-2}{(n-3)}\left[(n-3)(n-1)+2(n-2)be^{az/(n-3)}
  +b^2e^{2az/(n-3)}\right].
  \label{p}
 \en
 It is easy to see that we will get a negative density for a
 positive $b$.

 All the variables $t,z,x_1,...,x_{n-2}$ in (\ref{metric}) can run
 from $-\infty$ to $\infty$, thus the topology of this solution is
 simply $R^n$. But real singularity will appear if we do not
 constrain the permitted interval of $z$, which we will discuss in
 detail in Section IV.

      It would be useful to see different degenerations when the parameters take
      some special values before investigating the detailed properties of the full metric. Obviously when
 $a=0$ (\ref{metric}) degenerates to Minkowski metric. When $b=0$, it
 becomes
 \be
 ds^2=dz^2+e^{2az}(-dt^2+d\Sigma^2),
 \en
 ,
 which is just n-dimensional anti-de Sitter (AdS)
 metric.

 \section{Energy conditions}
    In some sense, any metric is a solution of Einstein field equation if we just treat the
    resulted Einstein tensor as the energy momentum tensor (up to a constant factor).
    The Einstein equation itself is not choosy to energy momentum form. However, when the gravitational
    physics is associated
     with other parts of physics (it must be), the form of energy momentum is constrained by the properties
     of the matter source. Omitting the specifics of different matters, people put forward several energy conditions
     for the matters. The ordinary matters obey all these conditions. But in some cases, we need an exotic matter.
     For example, due to the recent observation we need an unknown exotic matter to drive the acceleration of the universe.
     In the case of a Taub solution, we need a source with the exotic matter which violates DEC.   Now we begin to
   study the constraints by weak, strong and dominant energy
   conditions for the metric (\ref{metric}). In this section we consider a planar
   source with finite source, which inhabits in the region $z\geq
   0$.
   Our solution is an interior solution of plane source, hence pressure
   $p(z)$ should vanish naturally at some distance from the ``ground",
   $z=z_0 $, that is, we should match it to a vacuum solution. This
   condition means,
 \be
 p(z_0)=\frac{a^2}{2}\frac{n-2}{(n-3)}\left[(n-3)(n-1)+2(n-2)be^{az_0/(n-3)}
  +b^2e^{2az_0/(n-3)}\right]=0,
 \en
 which has two roots,
 \bea
 b_1=-(n-3)e^{-\frac{az_0}{n-3}},\\
 b_2=-(n-1)e^{-\frac{az_0}{n-3}}.
 \ena
 We call the solutions with $b_1$ and $b_2$ ``little branch" and ``large
 branch", respectively.

   For a metric like (\ref{metric}), the physical meanings of the parameters are not evident. A convenient way to
   show the meanings is to calculate the mass per area of this slab
   $\alpha$,
  \be
  \alpha=\int_0^{z_0}  dz e^{(n-2)\left[az+be^{az/(n-3)}\right]} \rho(z),
  \en
  where $\rho(z)$ is defined in (\ref{rho}). For the little branch $b=-(n-3)e^{-az_0/(n-3)}$,
  the integral yields a recurrence representation by integration in
  part. For example, in the case $n=5$,
  \bea
  \alpha=\frac{a}{1944}e^{-az_0/2-6e^{-az_0/2}}\left[3888-4212e^{az_0/2}-324e^{{-az_0}}
  -270e^{3az_0/2}-180e^{2az_0}-90e^{5az_0/2}-30e^{3az_0}\right.\\
  \left.-5e^{7az_0/2}+1233e^{-6-7az_0/2-6e^{az_0/2}}\right].
  \label{aremas1}
  \ena
  For the large branch, the result is
 \bea
  \alpha=\frac{a}{124416}e^{-az_0/2-12e^{-az_0/2}}\left[497664-269568e^{az_0/2}-10368e^{{-az_0}}
  -4320e^{3az_0/2}-1440e^{2az_0}-360e^{5az_0/2}\right.\\
  \left.-60e^{3az_0}-5e^{7az_0/2}+211543e^{-6-7az_0/2-6e^{az_0/2}}\right].
  \label{aremas2}
  \ena
  For $n>5$, the representations are much more complicated, but keep
  finite. Since the  mass per area is proportional to $a$,  $a$ can
  be treated as a mass parameter of the slab. Whether $\alpha$ is negative
  depends not only on $a$, but also on $z_0$. More generally, different energy conditions give different
  constraints on $az_0$.

 In the following subsections, we will study the constraints on $az_0$ by
 different energy conditions.

 \subsection{Weak energy condition}
 Weak energy condition (WEC) presents the constraint
 \be
 T(Z,Z)\geq 0,
 \en
 for any time-like vector $Z$. In a comoving frame of the fluids the above equation yields,
 \bea
 \rho=-\frac{a^2}{2}\frac{n-2}{(n-3)^2}\left[(n-3)^2(n-1)+2(n-2)^2be^{az/(n-3)}
  +(n-1)b^2e^{2az/(n-3)}\right]\geq 0,
 \label{1stwec} \\
 \rho+p=-\frac{a^2(n-2)}{(n-3)^2}\left[ (n-2)b
 e^{{az}/{(n-3)}} +  b^2e^{{2az}/{(n-3)}}\right]\geq 0
 \label{2ndwec}.
 \ena

 First, we consider the little branch $b=-(n-3)e^{-az_0/(n-3)}$. We only need to replace $b$ by $-(n-3)e^{-az_0/(n-3)}$
 in (\ref{1stwec}) and (\ref{2ndwec}). For convenience we
 investigate three cases $a>0$, $a<0$, $a=0$, respectively.

 Case I:
 $a>0.$
  Condition
 (\ref{1stwec}) becomes
 \be
 \frac{(n-2)^2-\sqrt{2(n-2)^2-1}}{(n-1)(n-3)}e^{az_0/(n-3)}
 \leq e^{az/(n-3)}\leq
 \frac{(n-2)^2+\sqrt{2(n-2)^2-1}}{(n-1)(n-3)}e^{az_0/(n-3)}.
 \label{wecc1}
 \en
 In the above equation the second sign of inequality is satisfied
 automatically since
 \be
 (n-2)^2+\sqrt{2(n-2)^2-1}-(n-1)(n-3)=1+\sqrt{2(n-2)^2-1}>0,
 \en
 and the first sign of inequality implies,
  \be
 e^{az_0/(n-3)}\leq\frac{(n-1)(n-3)}{(n-2)^2-\sqrt{2(n-2)^2-1}}
 .
 \en

 Inequality (\ref{2ndwec}) yields,

 \be
  \frac{n-2}{n-3}e^{az_0/(n-3)}\geq e^{az/(n-3)}\geq 0.
  \label{wecc2}
  \en
  The first sign of inequality of the above equation is satisfied
  automatically, since $z_0\geq z$ and $(n-2)/(n-3)>1$, and
  obviously the second sign of inequality is also satisfied
  automatically. Hence (\ref{2ndwec}) impose no constraint on the
  parameters in this case.

 Case II:
 $a<0.$ Again inequality
 (\ref{1stwec}) requires (\ref{wecc1}).
  Here, different from the case of $a>0$, the first sign of inequality is satisfied automatically since
 \be
 (n-2)^2-\sqrt{2(n-2)^2-1}-(n-1)(n-3)=1-\sqrt{2(n-2)^2-1}<0,
 \en
 and the second sign of inequality implies
 \be
 e^{az_0/(n-3)}\geq\frac{(n-1)(n-3)}{(n-2)^2+\sqrt{2(n-2)^2-1}}.
 \en
 In this case inequality
 (\ref{2ndwec}) still requires (\ref{wecc2}).
   But here,  it is not satisfied
  automatically. Especially, the first sign of inequality implies,
  \be
  e^{az_0/(n-3)}\geq \frac{n-3}{n-2}.
  \en
  The second sign of inequality is satisfied automatically.

 Therefore the permitted interval of $a$ and $z_0$ is
 \be
 e^{az_0/(n-3)}\geq \frac{n-3}{n-2},
 \en
 for the case $a<0$.
 The case $a=0$ degenerates to Minkowski space, which is a trivial
 case  satisfying any energy conditions. In summary, in the interval,
  \be
  e^{az_0/(n-3)}\in \left[\frac{n-3}{n-2}, ~\frac{(n-1)(n-3)}{(n-2)^2-\sqrt{2(n-2)^2-1}} \right].
  \label{wec1}
  \en
  WEC always can be satisfied for any real $a$ in the little branch.

  Second we consider the large branch. We then replace $b$ by $-(n-1)e^{-az_0/(n-3)}$
 in (\ref{1stwec}) and (\ref{2ndwec}). Similarly, we
 investigate three cases according to the sign of $a$.

 Case I:
 $a>0.$
  Inequality
 (\ref{1stwec}) requires
 \be
 \frac{(n-2)^2-\sqrt{2(n-2)^2-1}}{(n-1)^2}e^{az_0/(n-3)}
 \leq e^{az/(n-3)}\leq
 \frac{(n-2)^2+\sqrt{2(n-2)^2-1}}{(n-1)^2}e^{az_0/(n-3)}.
 \label{wecc1b}
 \en
  In the above equation the second sign of inequality  cannot   be satisfied in the neighborhood of $z=z_0$, since
 \be
 (n-2)^2+\sqrt{2(n-2)^2-1}-(n-1)^2=3+\sqrt{2(n-2)^2-1}-2n<0,
 \en
  while the first sign of inequality is implied by,
  \be
 e^{az_0/(n-3)}\leq\frac{(n-1)^2}{(n-2)^2-\sqrt{2(n-2)^2-1}}.
 \en
 In this case inequality
 (\ref{2ndwec})  requires
 \be
  \frac{n-2}{n-1}e^{az_0/(n-3)}\geq e^{az/(n-3)}\geq 0.
  \label{wecc2b}
  \en
 In the above equation, the first sign of inequality can not be satisfied in the
  neighborhood of  $z=z_0$. So, the permitted interval for $az_0$ in
  the large branch for WEC is an empty set.

 Case II:
 $a<0.$  Still, inequality
 (\ref{1stwec}) requires (\ref{wecc1b}).

 Here, the first sign of inequality is satisfied automatically since
 \be
 (n-2)^2-\sqrt{2(n-2)^2-1}-(n-1)^2=3-2n-\sqrt{2(n-2)^2-1}<0,
 \en
 and the second sign of inequality implies
 \be
 e^{az_0/(n-3)}\geq\frac{(n-1)^2}{(n-2)^2+\sqrt{2(n-2)^2-1}},
 \en
 which can never be satisfied.
 And
 (\ref{2ndwec})  requires (\ref{wecc2b}).
   Similarly, the first sign of inequality can never be satisfied in the
  interval $z\in [0,z_0]$.

 The case $a=0$ degenerates to Minkowski space, which is a trivial
 case  satisfying any energy conditions. In summary, there is no
 proper interval in which WEC can be satisfied in the large branch for non-trivial case.

 \subsection{Strong energy condition}

 Now we turn to the strong energy condition (SEC), which requires
 \be
 Ric(Z,Z)\geq 0,
 \label{str}
 \en
 where $Ric$ denotes the Ricci tensor of metric (\ref{metric}), and
 $Z$ is an arbitrary time-like vector. In a comoving frame of the fluid the condition (\ref{str})
 becomes,
  \bea
 \rho+(n-1)p=\frac{a^2}{2}\frac{n-2}{(n-3)^2}\left[(n-3)^2(n-1)(n-2)+2(n-2)(n^2-5n+5)be^{az/(n-3)}
 \right.
 \nonumber
  \\ \left.+(n-1)(n-4)b^2e^{2az/(n-3)}\right]\geq 0,
 \label{1stsec} \\
 \rho+p=-\frac{a^2(n-2)}{(n-3)^2}\left[ (n-2)b
 e^{{az}/{(n-3)}} +  b^2e^{{2az}/{(n-3)}}\right]\geq 0
 \label{2ndsec}.
 \ena

 First, similarly, we consider the little branch $b=-(n-3)e^{-az_0/(n-3)}$. Also, we investigate
 the three cases  $a>0$, $a<0$ and $a=0$, respectively. Since the case
 $a=0$ is just Minkowski, which satisfies any energy conditions, we
 omit it in the following discussions.

 Case I:
 $a>0$. The inequality (\ref{2ndsec}) has been discussed in the
 case WEC, which is satisfied naturally. One finds (\ref{1stsec})
 implies,

 \be
 e^{az/(n-3)}\geq
 \frac{(n-2)(n^2-5n+5)+\sqrt{(n-2)(n^3-8n^2+20n-14)}}
 {(n-1)(n-3)(n-4)}e^{az_0/(n-3)},
 \label{1stsecda}
 \en
 or
 \be
 e^{az/(n-3)}\leq
 \frac{(n-2)(n^2-5n+5)-\sqrt{(n-2)(n^3-8n^2+20n-14)}}
 {(n-1)(n-3)(n-4)}e^{az_0/(n-3)}.
 \label{2ndsecxi}
 \en
 Note that the case of $n=4$ in the above condition can not be
 obtained simply by direct replacing $n$ with 4 because the expression
 will diverge. The case for $n=4$ should be calculated separately
 and  it has been done in \cite{self}. The result is that SEC presents no
 constraints on the parameters for 4-dimensional solution.
 In the general case with $n>4$, inequality (\ref{1stsecda}) can not be
 satisfied since
 \be
 \frac{(n-2)(n^2-5n+5)+\sqrt{(n-2)(n^3-8n^2+20n-14)}}
 {(n-1)(n-3)(n-4)}>1,
 \en
 and (\ref{2ndsecxi}) is satisfied automatically since,
 \be
  \frac{(n-2)(n^2-5n+5)-\sqrt{(n-2)(n^3-8n^2+20n-14)}}
 {(n-1)(n-3)(n-4)}>1.
 \en
  So, SEC imposes no constraint on the
 parameters in this case.

 Case II:
 $a<0$. Inequality
 (\ref{2ndsec}) has been analyzed before, which requires
 \be
  e^{az_0/(n-3)}\geq \frac{n-3}{n-2}.
  \en
  Inequality
 (\ref{1stsec}) requires (\ref{1stsecda}) or (\ref{2ndsecxi}).
 Similarly to the case of $a>0$, the two inequalities can not
 be simply applied to the 4-dimensional solution, which is discussed
 in \cite{self}. There is no solution for (\ref{1stsecda}) and the
 solution for (\ref{2ndsecxi}) is
 \be
 e^{az_0/(n-3)}\geq
 \frac{(n-1)(n-3)(n-4)}{(n-2)(n^2-5n+5)-\sqrt{(n-2)(n^3-8n^2+20n-14)}}
 .
 \en

  To summarize the requirement
 of SEC on parameters $a$ and $z_0$ is given as,
  \be
  e^{az_0/(n-3)}\in[\frac{(n-1)(n-3)(n-4)}{(n-2)(n^2-5n+5)-\sqrt{(n-2)(n^3-8n^2+20n-14)}}, \infty).
  \label{sec}
  \en

  Second, we consider the large branch $b=-(n-1)e^{-az_0/(n-3)}$ for SEC. Following the above
  discussions we investigate
 the three cases according to the sign of $a$, respectively.

 Case I:
 $a>0$. The inequality (\ref{2ndsec}) gives
  \be
  \frac{n-2}{n-1}e^{az_0/(n-3)}\geq e^{az/(n-3)}\geq 0.
  \label{a>02ndsec}
  \en
 Here, the first sign of inequality can not be satisfied in the
  whole interval $z\in [0,z_0]$.

 Case II:
 $a<0$.

 Inequality
 (\ref{a>02ndsec}) can not be satisfied either.
  So, for large branch SEC is violated for any $a$ and $z_0$ in the interval
  $z\in [0,z_0]$.

  \subsection{Dominant energy condition}

 DEC is also a very important energy condition, which requires
 \be
 \rho\geq |p|.
 \en
 To remove the calculation of absolute value, we consider the cases of $p<0$ and $p\geq 0$, respectively.
 First, $p<0$ requires that
 \be
 \left[b+(n-1)e^{-az_0}\right]\left[b+(n-3)e^{-az_0}\right]<0,
 \en
 whose solution is not available, since the $p$ always reaches zero at
 $z=z_0$.
 Second, for the case $p\geq 0$, DEC implies
 \be
  -\frac{a^2(n-2)}{(n-3)^2}\left[(n-3)^3(n-1)+(2n-5)(n-2)be^{az/(n-3)}+(n-2)b^2e^{2az/(n-3)} \right]\geq 0.
  \label{dec1}
 \en
 From the result of the 4-dimensional case we learn that the large
 branch can perfectly match to Taub solution and the main underlying factor is
 the violation of the DEC. Hence
 here we emphatically investigate the requirements of DEC in the large branch.

 In the large branch (\ref{dec1}) becomes
 \be
 \frac{2n-5-\sqrt{\frac{5n-14}{n-2}}}{2(n-1)}e^{az_0/(n-3)}<e^{az/(n-3)}<
 \frac{2n-5+\sqrt{\frac{5n-14}{n-2}}}{2(n-1)}e^{az_0/(n-3)}.
 \en

 When $a>0$ the second sign of the above inequality is always violated  in the
 neighborhood of $z=z_0$ since
 \be
  \frac{2n-5+\sqrt{\frac{5n-14}{n-2}}}{2(n-1)}<1.
  \en
  This also leads to the violation of DEC in the neighborhood of
  $z=0$ when $a<0$. So, the matter of this source always violates DEC
  for the large branch, which is a necessary condition for the source of Taub solution.
  There is a typographical error in (36) in \cite{self}, that is, $\geq$ should
  be $\leq$.

  In table I we show a summary of the permitted regions of $e^{az_0/(n-3)}$ by different energy
  conditions. $LIB$, $LAB$, $ES$ denotes the little branch,
  the large branch, and
  empty set, respectively.
  {\huge
 \begin{table}
 \begin{center}
 \begin{tabular}{ccccc}
 \hline\hline
 $~~~~~~~$ &  ~~~~~~$WEC$&~~~~$SEC$&~~~~~$DEC$\\ \hline
 $LIB~a>0$ & ~~~~~$(0,\frac{(n-1)(n-3)}{(n-2)^2-\sqrt{2(n-2)^2-1}}]$ & $(0, \infty)$\\ \hline
 $LIB~a<0$ &~~~~~ $[\frac{n-3}{n-2}, ~\frac{(n-1)(n-3)}{(n-2)^2-\sqrt{2(n-2)^2-1}}]$ & $[\frac{(n-1)(n-3)(n-4)}{(n-2)(n^2-5n+5)-\sqrt{(n-2)(n^3-8n^2+20n-14)}}, \infty)$\\
 \hline
 $LAB~a>0$ & ~~~~~$ES$ & $ES$& $ES$\\ \hline
 $LAB~a<0$ &~~~~~ $ES$ & $ES$& $ES$\\ \hline
 $a=0~{\rm in~either~branch}$ &~~~~~ $(0, \infty)$ & $(0, \infty)$ & $(0, \infty)$\\
 \hline\hline
 \end{tabular}
 \end{center}
 \caption{\label{allow} Allowed regions of $e^{az_0/(n-3)}$ by different energy conditions.
 We have
  considered $e^{az_0/(n-3)}>0$ for any real $az_0$, which is not a constraint required
  by the
  energy conditions.}
 \end{table}}

 \section{matching to vacuum solutions}

 Though (\ref{metric}) is a rigorous solution
 with perfect fluid source, which can be filled in the whole space, serious
 problems will appear if it is really filled in the whole space.
 For example, there is a true singularity if interval of $z$ is not
 confined. This violates our original intention to study this
 solution: we aim to remove the time-like singularity in Taub space by
 replacing the singularity with matter source.
 In fact, the Ricci scalar $R$ reads,
  \be
  R=-\frac{2a^2}{(n-3)^2}\left\{\frac{1}{2}n(n-1)(n-3)^2+(n-2)[n(n-3)+1]be^{az/(n-3)}
  +\frac{1}{2}(n-2)(n-1)b^2e^{2az/(n-3)}\right\}.
  \en
 When $z$ goes to $\infty$ ($-\infty$), Ricci scalar will be divergent for
 a positive (negative) $a$. We hereby consider the case that this
 solution is only valid in a finite region and the spacetime is
 vacuum out of this region.

 The gravitational field must satisfy
 two boundary conditions: 1. The metric is continuous across the
 boundary surface, and 2. The extrinsic curvatures measured by the
 different sides of the boundary surface relate to each other by
 \be
 [K-h{\rm tr}(K)]^{\pm}=- \tau,
 \label{jump}
 \en
 in which $K$ denotes the extrinsic curvature of the boundary,
 $h=g-\frac{\partial}{\partial z}\otimes \frac{\partial}{\partial z}$
  represents the induced metric on the boundary,
 $\tau$ is the energy-momentum tensor confined to the boundary, and
 $[~]^{\pm}$ denotes the jump at the boundary, i.e., for a quantity
 $Q$, $[Q]^{\pm}=\lim_{{(z-z_0)} \to 0^{+}}Q(z)-\lim_{(z-z_0) \to
 0^{-}}Q(z)$.
 The exterior vacuum space has to be plane symmetric to match to
 metric (\ref{metric}). Although our original purpose is to
 find a source of Taub space, one may be curious about the case of the matching of this metric to Minkowski.
    So,  first we study the simplest case in which the vacuum out of the source region is Minkowskian
  geometry. Then we analyse the junction condition matching to the well known
  non-flat plane symmetric space: (static) Taub space.

  Before discussing the boundary condition between the slab and the vacuum, we
  impose a mirror boundary condition at $z=0$. Then the topology of the solution
  (\ref{metric}) becomes $R^{n-1}\times R/Z_2$, which means that we only need to study the
  region $z\geq 0$ of the resulting space. The continuity condition is naturally
  satisfied and the jump condition requires
  \be
  [K-h{\rm tr}(K)]^{0+}=-\frac{1}{2} (\tau) ^{0},
 \en
 where $0+$ labels the value of a quantity at $z=0$ when going
  from the positive direction, $0$ denote the value of a quantity
  at $z=0$. Using the above equation we derive
  \be
  \tau^{\nu}_{\mu}|_{z=0}=2 {\rm diag}((n-2)(a+\frac{ab}{n-3},(n-2)a+ab,...,(n-2)a+ab),
  \en
  where $...$ represents n-4 ((n-2)a+ab)s, and $b=-(n-3)e^{-az_0/(n-3)}$ or $b=-(n-1)e^{-az_0/(n-3)}$, depending on the little
  or large branch.

      We take the Minkowskian metric in the
  following chart,
  \be
  ds_{Min}=-l^2dt^2+dz^2+m^2(d\Sigma ^2),
  \label{min}
  \en
 where $l$ and $m$ are positive constants. The continuous condition yields
 \bea
 l&=&e^{az_0},\\
 m&=&e^{az_0+be^{az_0/(n-3)}},
 \ena
 where $l$ and $m$ denote ``time contraction'' and ``length
 contraction'', respectively
 though $l,~m$ may be greater than 1 so
 that they represent ``time dilation'' and ``length expansion''.
   From the jump condition (\ref{jump}) we obtain
 $\tau$ in induced chart by (\ref{min}) in the little branch,
 \be
 \tau_{~\mu}^{\nu}|_{z=z_0}={\rm diag}({0,~a,...,~a}),
 \label{mmin1}
 \en
 while in the large branch,
 \be
 \tau_{~\mu}^{\nu}|_{z=z_0}={\rm diag}({-2(n-2)a,-a,...,-a}).
 \label{mmin2}
 \en
 When $a=0$, $\tau$ vanishes, which is just our expectation because
 the interior metric degenerates to Minkowski when $a=0$. Note that
 the boundary energy-momentum is determined up to a sign, that is, depends
 on which side, the vacuum or the source, is regarded as the
 positive direction.

 Next we discuss the conditions matching to the n-dimensional Taub metric. The n-dimensional Taub metric
 reads,
 \be
 ds^2=-z^{2\alpha}k^2dt^2+dz^2
 + z^{2\beta}l^2d\Sigma^2,
 \label{ntaub}
 \en
 where $\alpha=-\frac{n-3}{n-1},~\beta=\frac{2}{n-1}$. $k, l$ are two constants.
 The continuous condition requires
 \bea
 k=\pm \frac{e^{az_0}}{z_0^{\alpha}},
 \label{azl}
 \ena
  and
 \be
 l=\pm \frac{e^{az_0+be^{az_0/(n-3)}}}{z_0^{\beta}}.
 \label{azb}
 \en
   The jump condition (\ref{jump}) gives,
 \bea
 \tau_{~\mu}^{\nu}|_{z=z_0}={\rm diag}\left(-(n-2)\beta
 z_0^{-1}+(n-2)(a+\frac{ab}{n-3}e^{az_0/(n-3)}),~
 ((3-n)\beta-\alpha)z_0^{-1}+(n-2)a+abe^{az/(n-3)}\right. \nonumber \\
 \left.  ...,
 ((3-n)\beta-\alpha)z_0^{-1}+(n-2)a+abe^{az/(n-3)}\right).
 \ena
  For the little branch, it does not vanish for any finite
  parameters. For large branch, it becomes,
 \be
 \tau_{~\mu}^{\nu}|_{z=z_0}=(a+\frac{n-3}{n-1}z_0^{-1}){\rm
 diag}\left(-\frac{2(n-2)}{n-3},~-1,...,-1\right).
 \en
 We see that the matter filled in the boundary is quintessence-like, with constant
 equation of state (EOS) $-\frac{n-3}{2(n-2)}$. It is well-known that in cosmology such an energy-momentum can be
 simulated by a scalar field with well connected potential. If we
 require the matching is perfect, that is, the boundary
 energy-momentum vanishes, we obtain,
   \be
    az_0=-\frac{n-3}{n-1}.
    \en
 Therefore, the solution (\ref{metric}) is reasonably treated as the source of
 n-dimensional Taub space (\ref{ntaub}). Here we correct an error in
 \cite{self}, which we also pointed out in \cite{self2}. The correct
 form of (68) in  \cite{self} should be
 \be
 \tau_{~\mu}^{\nu}=(a+\frac{1}{3}z_0^{-1}){\rm
 diag}\left(-4,~-1,-1\right).
 \en
   So the condition for perfect matching to the vacuum space becomes
   $az_0=-1/3$.

\section{Conclusions and discussions}
 In this paper we present an n-dimensional global solution of Einstein field equation
  with a perfect fluid source,
 which we interpret as the source of some planar symmetric vacuum space.
 $(n-2)(n-1)/2+1$ Killing vectors, including a time-like Killing vector,
 are permitted in this solution. We find a chart in which the metric is
 written in time coordinate orthogonal form.

 We find the ranges of the parameters in which WEC and SEC can be
 satisfied, respectively. Interestingly, we find that DEC is always
 violated, no matter what values the parameters are taken in the large branch, which is a
 necessary condition for the source of Taub solution.

  We do some preliminary researches on matching to vacuum solutions.
 Minkowski's and Taub's solution are studied respectively. The boundary energy-momentum never
 vanishes for matching to Minkowski. On the contrary, there is a special configuration
 in the present solution which can perfectly match to Taub space.

 {\bf Acknowledgments.}
   H.Noh was supported by grant No. C00022 from the Korea Research
  Foundation.

\end{document}